\documentstyle[12pt]{article}
\normalsize
\def\dlta{\bigtriangleup}

\def\be{\begin{equation}}
\def\ee{\end{equation}}
\def\bea{\begin{eqnarray}}
\def\eea{\end{eqnarray}}

\newcounter{sxn}

\newcounter{axn}

\def\br{\begin{thebibliography}{abc}}
\def\er{\end{thebibliography}}

\begin{document}
\bibliographystyle{unsrt}
\footskip 1.0cm
\thispagestyle{empty}
\begin{flushright}
IP-BBSR/95-83\\
Sept. 1995\\
\end{flushright}
\vspace*{10mm}
\centerline {\LARGE Formation of Topological Defects with}
\vspace*{3mm}
\centerline {\LARGE Explicit Symmetry Breaking}
\vspace*{8mm}
\centerline {\large Sanatan Digal and Ajit M. Srivastava}
\vspace*{5mm}
\centerline {\it Institute of Physics}
\centerline{\it Sachivalaya Marg, Bhubaneswar 751005}
\centerline {\it India}
\vspace*{8mm}

\baselineskip=16pt

\centerline {\bf ABSTRACT}
\vspace*{4mm}

  We demonstrate a novel mechanism for the formation of topological
defects in a first order phase transition for theories in the presence
of small explicit symmetry breaking terms. We carry out numerical
simulations of collisions of two bubbles in 2+1 dimensions for
a field theory where U(1) global symmetry is spontaneously as well
as explicitly broken. In the  coalesced region of bubble walls,
field oscillations result in the decay of the coalesced portion in a large
number of defects (e.g. ten  vortices and anti-vortices). We discuss
the implications of our results for axionic strings in the early
Universe, for baryon formation in quark-gluon plasma, and for
strings in liquid crystals in the presence of external
electric or magnetic field.

\newpage

 Production and subsequent evolution of topological defects has been
of considerable interest for particle physicists in the context
of the early Universe \cite{shlrd}. Similar techniques have also been
used to study baryon formation during hadronization of quark-gluon
plasma (QGP) in heavy ion collisions \cite{brn,kpst}. Study of
topological defects has, of course, been possible in most detailed way
only in condensed matter systems where they can be experimentally
studied \cite{nlc}.

 The aforementioned defects correspond to a spontaneous breakdown of a
symmetry. However, there are many situations
when the symmetry is also explicitly broken. In particle physics, the
Pecci-Quinn scheme for solving the strong-CP problem of quantum
chromodynamics leads to  the presence of an explicit symmetry
breaking term and, consequently, to axionic strings \cite{axn}.
The Skyrmion picture of baryons in the context of
chiral models is another example where explicit symmetry breaking
terms are needed to incorporate a non-zero pion mass \cite{kpst}. In condensed
matter, liquid crystals provide a simple example of such systems
where the presence of external electric or magnetic fields induces
explicit symmetry breaking terms \cite{lc}

  The study of formation of topological defects in such systems is therefore
important as it has implications for a diverse set of phenomena. It
has recently been argued that explicit symmetry breaking can lead to
a four fold enhancement in the production of baryons in QGP \cite{kpst}.
These arguments were largely qualitative and did not depend sensitively
on the order of the phase transition. It was argued in \cite{kpst} that
a similar enhancement should occur for other topological defects as well.

  In this letter we demonstrate a new mechanism for the production of
topological defects for systems with explicit symmetry breaking and
with a first order phase transition, where phase transition proceeds via
bubble nucleation. This mechanism leads to a
much stronger enhancement in defect production, and results from
a combination of the effects discussed in \cite{kpst} as well as
effects coming from the large field oscillations in the region of
coalesced bubble walls. The net result is that wall oscillations decay by
producing a large number of vortices and antivortices. For example in one
simulation, we  found 5 vortices and 5 antivortices produced
in a single two-bubble collision.

  We adopt the same numerical technique as used in previous simulations
of vortex formation via bubble collision, see \cite{ajt}. We will study
vortex formation in 2+1 dimensions in a field  theory system described
by the following Lagrangian

\begin{equation}
{\it L} = {1 \over 2} \partial_{\mu} \Phi^{\dag} \partial^{\mu} \Phi
- {\lambda \over 4} \phi^2 (\phi -\phi_0)^2 +
\epsilon \phi_0 \phi^3 + \kappa {\phi_0}^2 \phi^2 cos \theta
\end{equation}

\noindent where $\phi$ and $\theta$ are the magnitude and the phase of the
complex scalar field $\Phi$ ($\Phi = \phi e^{i \theta}$).
This Lagrangian describes a theory where the U(1) global symmetry is
spontaneously broken, except for the presence of the last term which
breaks this U(1) explicitly. When $\kappa$ is zero,
this theory allows for the existence of a cylindrically symmetric vortex
which is a  solution of the time independent field equations. For a
non-zero $\kappa$, the vortex looses azimuthal symmetry and is not
a solution of the time independent equations of motion any more.

  For $\kappa = 0$, the process of vortex creation via bubble nucleation
has been described in \cite{ajt}. At zero temperature, bubbles of
true vacuum nucleate via quantum tunneling in
the background of the metastable vacuum
with $\phi = 0$. These are described by the bounce solution which is
an O(3)-symmetric, least action, solution of the Euclidean field
equation \cite{bbl}

\be
{d^2 \phi \over dr^2} + {2 \over r} {d \phi \over dr} -
V^\prime(\phi) = 0
\ee

\noindent where $V(\phi)$ is the effective potential in Eq.(1)
(with $\kappa = 0$) and $r$ is
the radial coordinate in the Euclidean space. In the Minkowski space
the profile of the nucleated bubble is obtained from the solution
of Eq.(2) by putting $t$ = 0. This bubble then evolves according to the
classical field equations obtained from the Lagrangian in Eq.(1) in
Minkowski space,

\be
\Box \Phi_i = - {\partial V(\Phi) \over \partial \Phi_i},~i=1,2
\ee

\noindent where $\Phi = \Phi_1 + i \Phi_2,~\Box$ is the d'Alembertian and
time derivatives of fields are set equal to zero at $t = 0$. In a phase
transition, $\theta$ varies randomly from one bubble to another. These
bubbles expand and vortices form at the junction of three or more bubbles if
the phase $\theta$ traces a nontrivial winding in that region. This is the
conventional Kibble mechanism of defect formation \cite{kbl} which leads
to the probability of vortex formation for 2 space dimensions equal to
1/4 per bubble \cite{nlc}.

   We wish to study the case when $\kappa$ is non-zero. First we briefly
recall the physical picture described in \cite{kpst}.
Consider a two bubble collision with the phase $\theta$ in the two
bubbles taking values $\pi + \alpha$
and $\pi - \alpha$ where $\alpha$ is small. As the bubbles collide,
$\theta$ in the coalesced portion will assume value $\pi$ due to the
geodesic rule (essentially to minimize energy) and will keep evolving
towards zero inside the bubbles.
It is then easy to see that this leads to a winding one getting created
on one end of the coalesced wall and winding minus one on the other
end \cite{kpst}. It was argued in \cite{kpst} that this leads to roughly
four fold enhancement in the number density of vortex production per bubble.

 However, it turns out that the actual dynamics of
vortex creation has much richer structure, especially for
a first order phase transition. As
$\theta$ in both the bubbles evolves towards zero, the coalesced
portion of the walls undergoes large oscillations. Such oscillations have
been described in \cite{ajt} where it was shown (for the case
of subcritical bubbles) that, as $\phi$ undergoes large oscillations,
it passes through $\phi = 0$ forcing $\theta$ to change to $\theta + \pi$.

  This flip in the orientation of $\Phi$ has very important effects on
the process of vortex formation. The evolution of $\theta$ towards zero inside
the bubbles tends to create a winding one near one end and anti winding
near the other end inside the coalesced region. The flip in the
orientation of $\Phi$ in the central region completes
these windings and results in the nucleation of a vortex-antivortex
pair in the coalesced wall.

 This explains the formation of the first vortex-antivortex pair. Subsequent
pairs are created due to the oscillation of $\theta$ about $\theta = 0$.
As $\theta$ in the two bubbles evolves towards zero, it overshoots
the true vacuum (i.e. $\theta$ = 0). It is easy to convince oneself that
this evolution of $\theta$ in the two bubbles, combined with
the flipping of the orientation of $\Phi$ due to wall oscillations, will
result in the creation of another vortex-antivortex pair.
This process continues as long as $\theta$ and $\phi$ oscillate about
$\theta=0$ and $\phi = 0$ respectively.

It is important to realize that if the $\phi$ oscillations do not have large
enough amplitudes, and associated with it appropriate $\theta$ oscillations,
then vortices are not nucleated. We see this in our simulations where
many oscillations of the wall, and that of $\theta$, may pass by before a
given pair gets nucleated. Also, the sequence of vortices and anti-vortices
created on one end of the coalesced region is quite arbitrary, depending
on the details of $\phi$ and $\theta$ oscillations; though, the net winding
number is always zero. This implies that the annihilation of
vortex-antivortex pairs may be very ineffective.

We now proceed to describe our numerical results.
We find the bubble profile by solving Eq.(2) for $\kappa$ = 0 in $V(\phi)$.
This bubble profile will be an approximate solution of the
field equations obtained from Eq.(1) for small non-zero values of
$\kappa$ and provides an adequate starting point as bubbles collide only
after undergoing large expansions. We choose the values of the parameters
as, $\epsilon = 0.2,~ \phi_0 = 4.0,~ \lambda = 4.0$.
We use natural system of units with $\hbar = c = 1$ and measure
lengths in terms of $m_H^{-1}$, where $m_H$ is the Higgs mass in the
broken phase for the case when $\kappa$ is zero. $m_H^{-1} = 0.12$
for the above set of parameters. We have studied a range of values of
$\kappa$. Large values of $\kappa (> 0.12)$ do not give any vortex
formation as $\theta$ in bubbles rolls down and settles to zero before bubbles
can effectively coalesce. For all other values of $\kappa$ vortices
form (with smaller $\kappa$ leading to vortex formation at a later stage).
The figures shown in this paper correspond to the choice of
$\kappa = 0.06$.

Following the techniques developed in
\cite{ajt}, we study the case of a two bubble collision by prescribing
the nucleation centers for the two bubbles. This amounts to replacing
a portion of the false vacuum region (with $\Phi$ = 0) by the profiles of
two bubbles. The field configuration is then evolved by using a
discretized version of Eq.(3). Simulation is implemented by using a
stabilized leapfrog algorithm of second
order accuracy in both space and time. We use a 1000 $\times$ 1400
lattice with the lattice spacing in spatial directions, $\dlta x$,
equal to 0.104 (in the units of $m_H^{-1}$)
and lattice spacing in the temporal direction,
$\dlta t$, equal to $\dlta x / \sqrt{2}$. With these values, the evolution
was completely stable and energy was conserved within a few percent
during the simulation. For details of the numerical technique, see \cite{ajt}.
Simulations were carried out on a HP-735 workstation at the Institute
of Physics, Bhubaneswar.

 Bubble centers were chosen to lie at the y boundaries of the
lattice (and at the midpoint of the x axis)
so that the initial bubble profiles are that of half bubbles.  We
use free boundary conditions. Fig.1a shows the plot of $\Phi$ and the
contour plot of $\phi$ for the initial field configuration of one
of the bubbles. Values of $\theta$ for the two bubbles
are taken as 0.56 and 0.44 radians (for the lower and the upper
bubbles respectively) and are uniform inside each bubble.
This leads to development of a region of $\theta=\pi$ in the region
where bubbles coalesce. Fig.1b shows the plot of $\Phi$ at an intermediate
stage. The bubbles have significantly coalesced
and $\theta$ inside the bubbles has started rotating towards zero. The
rotation of $\theta$ is smaller near the bubble walls due to
the dependence of the explicit symmetry breaking term in Eq.(1) on $\phi^2$.
[This is the reason that, even if the initial values of $\theta$ in the
two bubbles are not close to $\pi$, a region of $\theta = \pi$ developes
in the coalesced region and vortices are produced.]

Fig.1c and Fig.1d show the configurations after the first pair has been
nucleated. The plot of $\Phi$ clearly shows that $\theta$ has overshot
the true vacuum ($\theta = 0$). Afterwards, $\theta$ starts climbing
towards $\pi$ first and then again rolls back towards zero. As described
earlier, this will cause creation of subsequent pairs for appropriate
$\phi$ oscillations. Fig.1e and
Fig.1f show the plots at a stage when there is a total of 10 vortices
and antivortices. Out of these, there are two groups, containing
three vortices each (as confirmed by detailed plots of $\Phi$ of
these regions) which are not well separated. Over all there are
at least 6 vortices and antivortices which are well separated.
Note that, due to the presence of explicit symmetry
breaking term, the profiles of these vortices are highly deformed as
shown by the contour plots.

 In conclusion, we have demonstrated a new mechanism for the
formation of topological defects in the presence of explicit symmetry
breaking which may dominate over other mechanisms of defect production.
A somewhat modified version of this mechanism (due to the absence
of coalesced portion of bubble walls) may also be applicable for the
case of second order phase transitions. [In this context we mention that
in \cite{ajt} it was found that the number of vortices produced was
roughly twice of the estimate based on the Kibble mechanism (though many
pairs annihilated quickly). In view of
our results in this paper, the excess production in \cite{ajt} suggests
that a nontrivial dynamics of $\theta$ coupled with $\phi$ oscillations
may contribute to vortex production even when explicit symmetry breaking
is absent.]  The most interesting aspect of this mechanism is that
it is literally a pair creation process, though still governed
by classical equations of motion. In this sense, it resembles
the pair creation of vortices in the flow of superfluid $^4$He
through a small orifice, as discussed in \cite{he4}, though actual
mechanisms are completely different. In a subsequent paper we will
present the study of full phase transition by nucleating large
number of bubbles \cite{dgl2}.

  Implications of these results are many. Using the ideas described above,
it is possible to argue that, in two bubble collisions in
3+1 dimensions, field oscillations should lead to string loops being
emitted out from the coalesced region. For axionic strings
in the early Universe, earlier studies have assumed that
the formation mechanism is the same as for other cosmic strings,
namely, via the Kibble mechanism \cite{axn}. Above discussion shows
that the dominant mechanism may be via the mechanism
discussed in this paper, at least for a first order phase transition.
Therefore, the final distribution of
axionic strings, and hence the frequency distribution of emitted axions,
can be drastically different from what is conventionally taken. Also,
as now one expects small string loops to be produced, axionic
domain walls may not survive for long. This may make a larger class of
axionic models viable.

  For the case of liquid crystals in the presence of electric field, our
results suggest that, instead of long strings, small string loops
should form in the coalesced region of two
bubbles. However, in this case, the dynamics
of string formation via this mechanism may be completely dominated by the
presence of damping terms.  The mechanism discussed in this
paper should also be applicable to the production of other defects,
though the details of the mechanism will depend on the type of defect
(and dimensionality of physical space) under consideration.
Especially important is the production of baryons (in the Skyrmion
picture) in quark-gluon
plasma \cite{kpst}. It is clearly important to investigate
the enhancement (which may be very large, as suggested by our results)
expected in baryon production due to explicit symmetry breaking terms
if the phase transition is of first order.

\vskip .3in
\centerline {\bf ACKNOWLEDGEMENTS}
\vskip .1in

 We would like to thank P. Agrawal and S.M. Bhattacharjee for useful
comments and B. Agrawal and P.A. Sriram for help in making figures.

\newpage

\vskip .3in
\centerline {\bf FIGURE CAPTIONS}
\vskip .1in

Fig.1: (a) Plot of $\Phi$ and contour plot of $\phi$ for the
initial configuration of one of the bubbles with center at the boundary.
For all $\Phi$ plots, the orientation of the arrows from positive x axis
gives the phase $\theta$ of $\Phi$ while the length of arrows is
proportional to $\phi$.
(b) Plot of $\Phi$ for the coalesced region at t = 55.2
showing that $\theta$ has significantly rotated towards zero in bubble
interiors. (c) Contour plot of $\phi$ at t = 62.6 showing a
vortex-antivortex
pair. (d) Winding numbers of the vortex and the antivortex
are clear from the $\Phi$ plot.
(e) Contour plot of $\phi$ at t = 84.7 showing
ten vortices and antivortices. There are two groups of three
overlapping vortices each, one near x = 34 and the other near x = 132.
These groups have net windings of $+1$ and $-1$ respectively.
(f) $\Phi$ plot  for a portion of lattice showing winding numbers of
at least 2 vortices and 1 antivortex which are well separated. The winding
$+1$ region near x = 34 actually consists of closeby configurations of
two vortices and one antivortex.

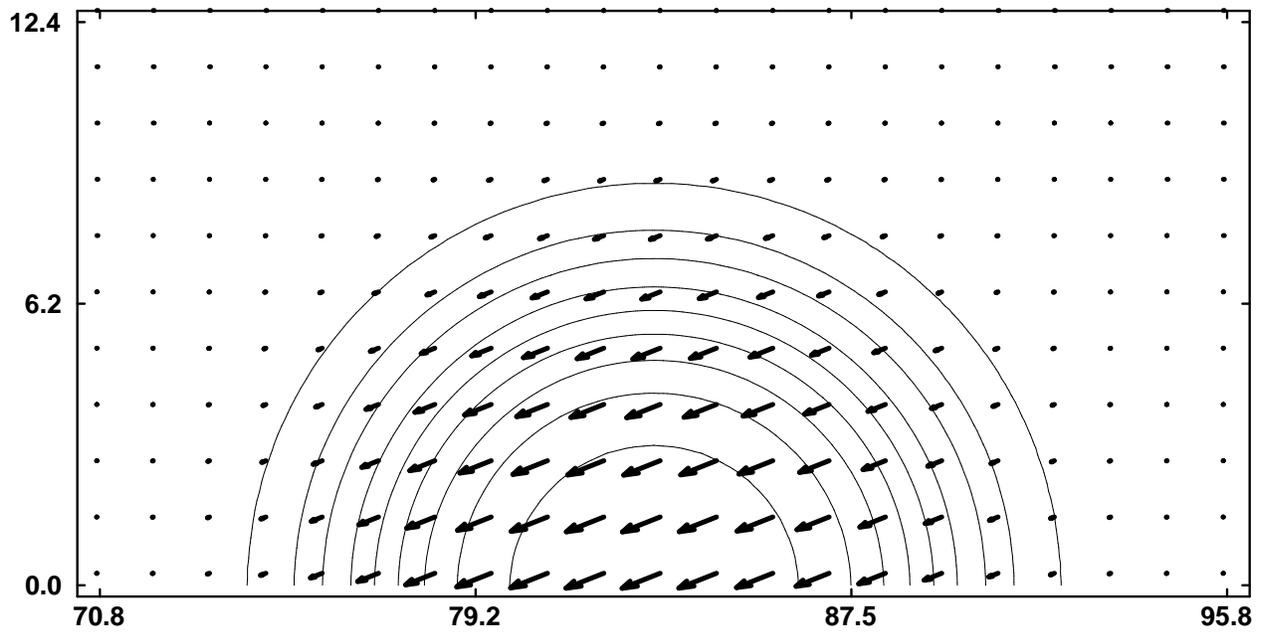
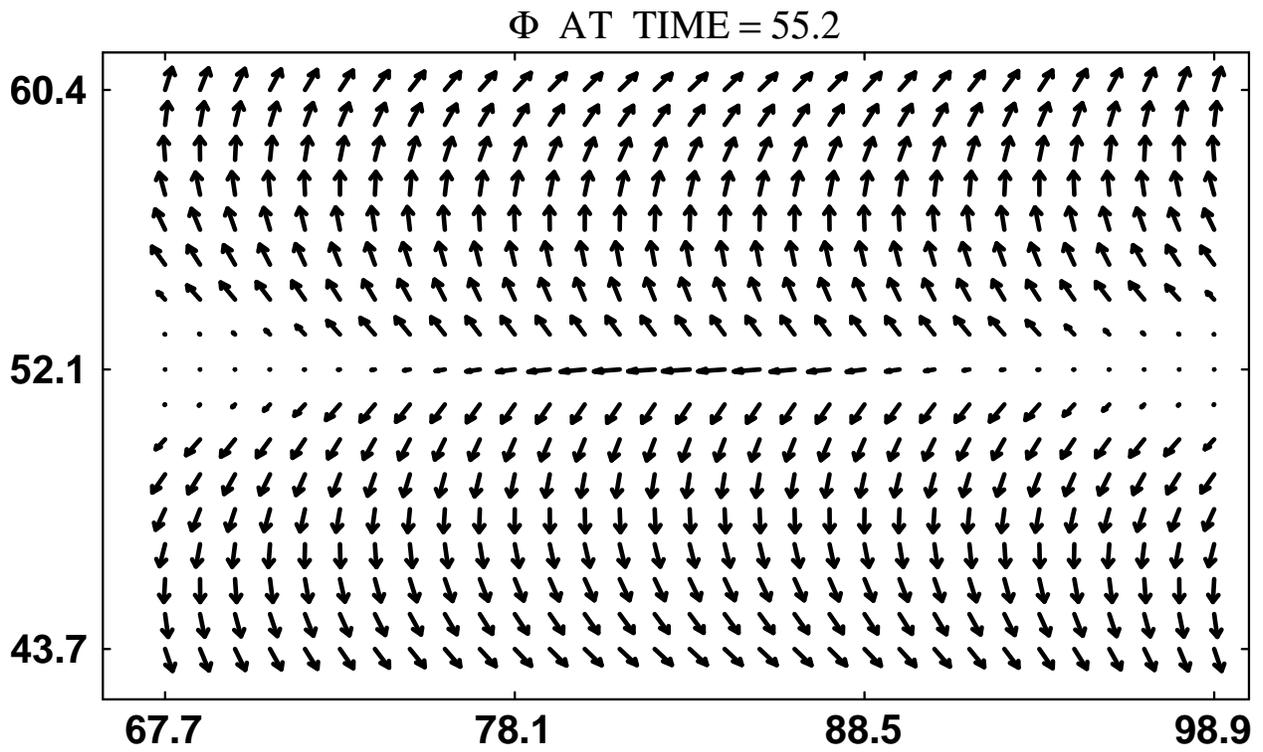
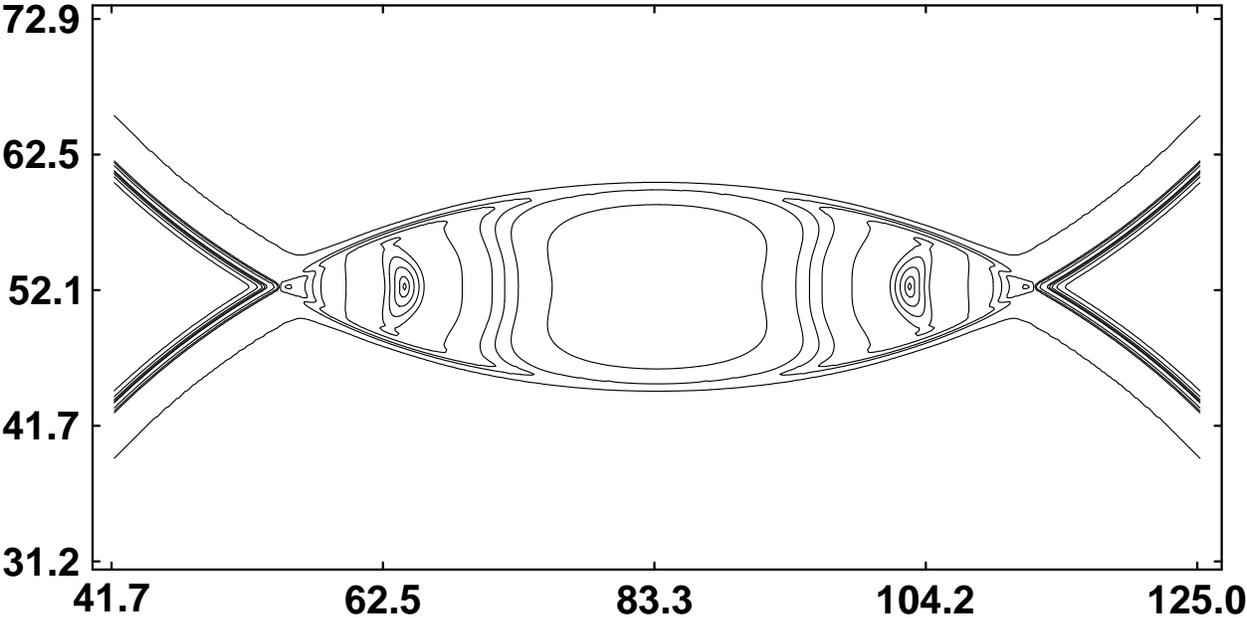
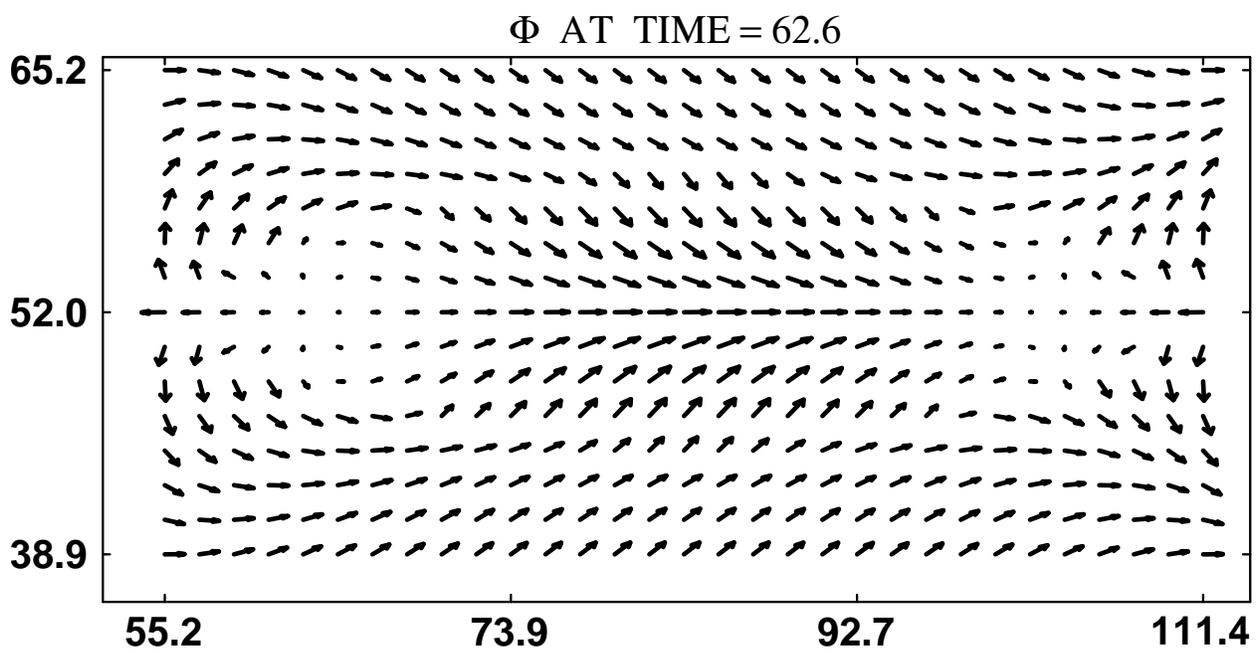
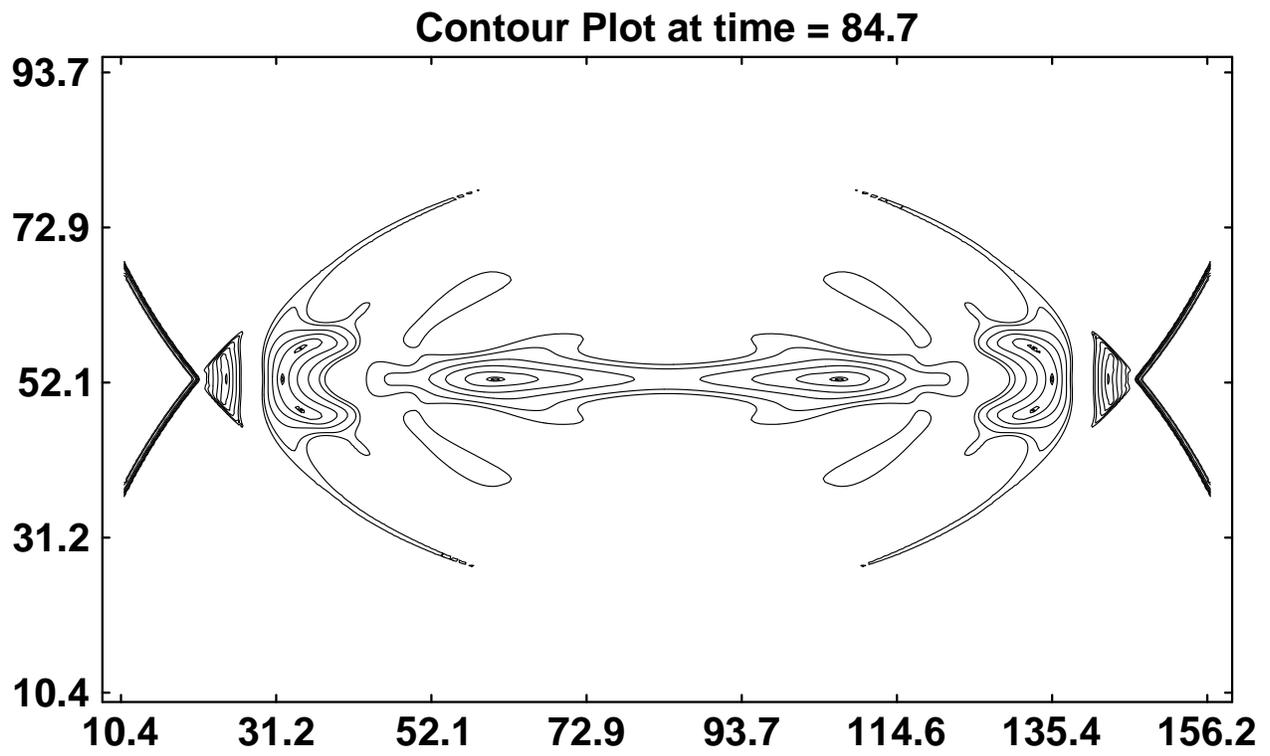
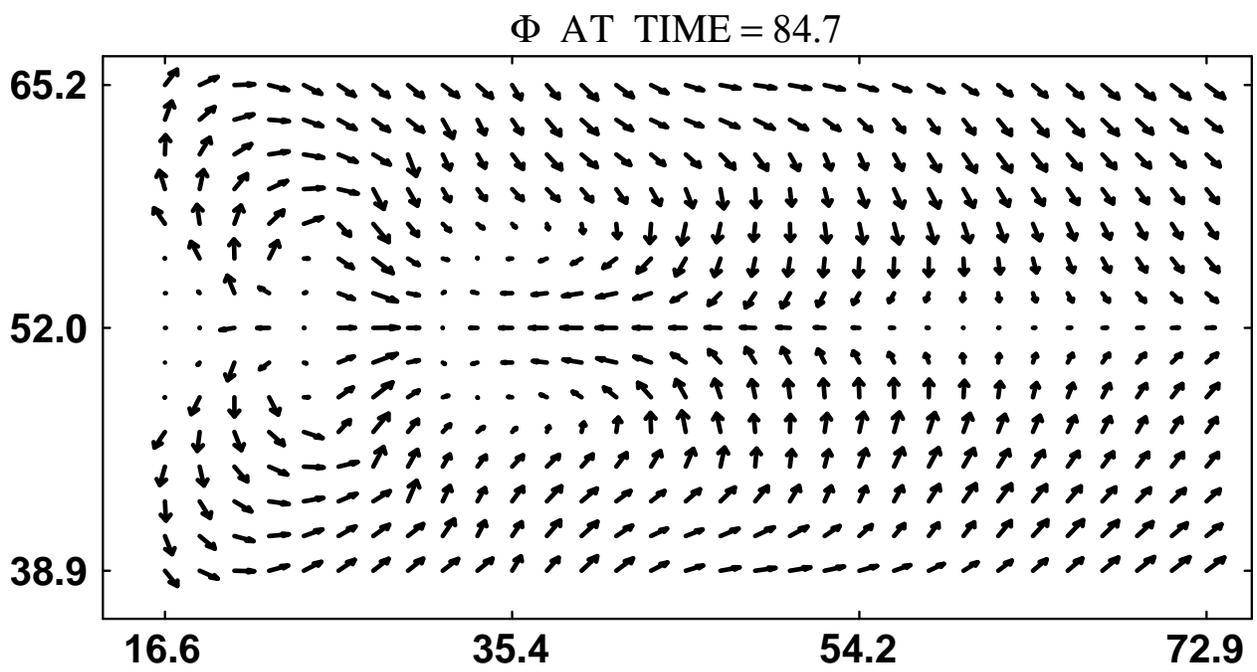
\end{document}